\documentclass[prl,10pt, twocolumn, nofootinbib, longbibliography]{revtex4-1}

\usepackage{graphicx,amsmath,amssymb}
\usepackage{color}
\usepackage{ulem}

\usepackage{lmodern}

\setcitestyle{super,longbibliography}

\def\Rb{$^{87}$Rb }

\def\ket#1{\left|#1\right\rangle}
\def\bra#1{\left\langle#1\right|}
\def\ketF#1{\left(#1\right)}

\def\ev#1{\left\langle#1\right\rangle}

\newcommand{\be}{\begin{equation}}
\newcommand{\beq}{\begin{eqnarray}}  
\newcommand{\ee}{\end{equation}}
\newcommand{\eeq}{\end{eqnarray}} 
\newcommand{\y}{{\rm YES}}
\newcommand{\n}{{\rm NO}}

\newcommand{\num}{\tilde{N}_{+1}}
\newcommand{\weightn}{\rho_n}
\newcommand{\weighto}{\rho_0}

\begin{document}
\hbadness = 10000

\title{Interaction-free measurements by quantum Zeno stabilisation of ultracold atoms}

\author{\begin{flushleft} \textsf{\textbf{J. Peise$^1$, B. L\"u{}cke$^1$, L.~Pezz\'e$^2$, F.~Deuretzbacher$^3$, W.~Ertmer$^1$, J. Arlt$^4$,  A.~Smerzi$^2$, L. Santos$^3$, C. Klempt$^{1*}$ 
}} \end{flushleft}}

\affiliation{\textsf{$^1$Institut f\"ur Quantenoptik, Leibniz Universit\"at Hannover, Welfengarten~1, D-30167~Hannover, Germany \\
$^2$Quantum Science and Technology in Arcetri (QSTAR), Istituto Nazionale di Ottica (INO), Consiglio Nazionale delle Ricerche (CNR), and European Laboratory for Non-Linear Spectroscopy (LENS), 50125 Firenze, Italy\\
$^3$Institut f\"ur Theoretische Physik, Leibniz Universit\"at Hannover, Appelstra\ss{}e~2, D-30167~Hannover, Germany
\\
$^4$QUANTOP, Institut for Fysik og Astronomi, Aarhus Universitet, 8000 \AA{}rhus C, Denmark\\
$^*$ Correspondence should be addressed to C.K. (email:klempt@iqo.uni-hannover.de)
}}

\maketitle
\textbf{
Quantum mechanics predicts that our physical reality is influenced by events that can potentially happen but factually do not occur. Interaction-free measurements (IFMs) exploit this counterintuitive influence to detect the presence of an object without requiring any interaction with it. Here we propose and realize an IFM concept based on an unstable many-particle system. In our experiments, we employ an ultracold gas in an unstable spin configuration which can undergo a rapid decay. The object - realized by a laser beam - prevents this decay due to the indirect quantum Zeno effect and thus, its presence can be detected without interacting with a single atom. Contrary to existing proposals, our IFM does not require single-particle sources and is only weakly affected by losses and decoherence. We demonstrate confidence levels of $90$\%, well beyond previous optical experiments.
}

After early work of Renninger~\cite{Renninger1953}, Elitzur and Vaidman~\cite{Elitzur1993} showed that the presence of an absorbing object in one arm of a Mach-Zehnder interferometer can be detected with a single photon, even if it passes through the other arm (see Fig.~1~a). In the absence of the object, constructive and destructive interference lead to a bright and a dark output port. However, if an object blocks the upper arm, the interference is absent and the photon can exit from the formerly dark output port - witnessing the existence of the object. In the literature, this has been termed an "interaction-free measurement" of the object, although the quantum mechanical description actually includes an interaction. The detection is only "interaction-free" when the photon leaves the dark output port, whereas a photon in the bright output port yields no information and a photon hitting the object corresponds to the case with interaction. 
Because of these unwanted results, the efficiency of interaction-free measurement is, at most, $50$\%~\cite{Elitzur1993,Vaidman2003}. It can be increased by exploiting the Zeno effect~\cite{Misra1977}, as proposed~\cite{Kwiat1995a} and experimentally verified~\cite{Kwiat1999} with polarized photons (see also Refs.~\citenum{Paul1996,Tsegaye1998} for an alternative proposal exploiting the resonance condition of a high-finesse cavity). 

\begin{figure}[t!]
		\includegraphics{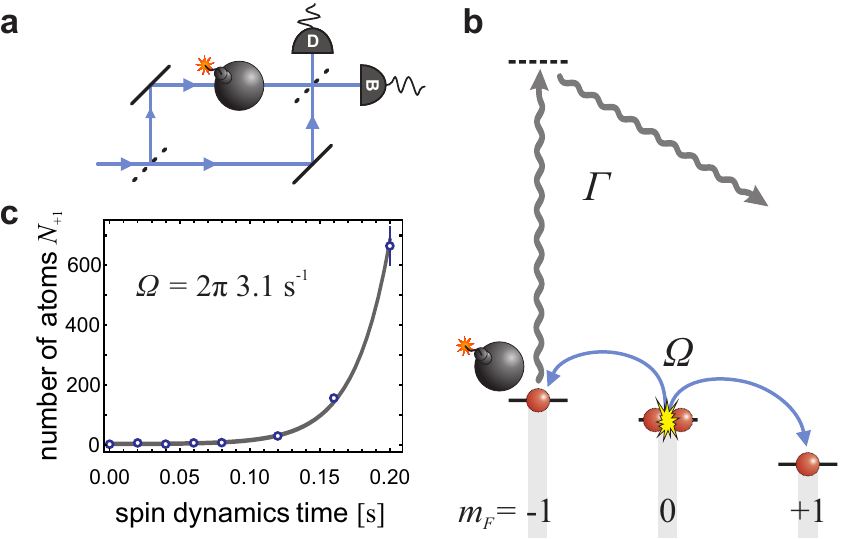}
		\caption{\textbf{Interaction-free measurements. a} Classic proposal for an interaction-free measurement~\cite{Elitzur1993}: A single photon entering the Mach-Zehnder interferometer never reaches the dark port D due to destructive interference. Only if an absorbing object (depicted as a bomb) is placed in one arm of the interferometer, the photon can trigger detector D, thereby witnessing the existence of the object. \textbf{b} Our implementation relies on a Bose-Einstein condensate in the Zeeman level $\ketF{F,m_F}=\ketF{1,0}$, which can decay to the levels $\ketF{1,\pm 1}$ by spin changing collisions. The analogue of an absorbing object is realized by a resonant laser beam, providing an effective loss of atoms in the level $\ketF{1,-1}$. Due to the quantum Zeno effect, this object will prevent the generation of atomic pairs by spin dynamics. An IFM of the object can thus be realized by counting the number of atoms in the level $\ketF{1,1}$: If the result is zero, the object exists and has not interacted with atoms. \textbf{c} Without the absorbing object, the number of atoms in level $\ketF{1,1}$ (open circles) grows exponentially according to $\ev{N_{+1}}=\sinh^2 (\Omega t)$ (solid line), which can be used as a calibration of the spin dynamics rate $\Omega$. The error bar presents the standard error of the mean number of transferred atoms.}
	\label{fig1}
\end{figure}

Here, we exploit the quantum Zeno effect to suppress the decay of an unstable system and use this principle for IFMs with an ideal efficiency of $100$\%. In the generic formulation of the quantum Zeno effect, an unstable system does not decay if its state is continuously measured. This continuous measurement can also be replaced by a continuous absorption of the decay products. An object that continuously absorbs the decay products therefore strongly suppresses the decay of the unstable system. In the limit of strong absorption, decay products are never generated. Therefore, the presence of the object can be detected interaction-free by monitoring whether the system decayed or not.

In our implementation, the system is realized by a Bose-Einstein condensate (BEC) in an unstable spin configuration. The decay products are atoms that are generated in pairs with opposite spin orientation. Hence, our experiments present a demonstration of the Zeno effect in a truly unstable many-body system~\cite{
[{For unstable single particles, see }][{}]Wilkinson1997,Fischer2001}. The "absorbing" object is realized by a resonant laser beam which removes the decayed atoms with one spin orientation from the system. The interaction-free character of the measurement is proven by a detection of the atoms with the opposite spin direction using a homodyne detection method~\cite{Gross2011}. Our implementation in principle allows for an arbitrarily high probability for an interaction-free measurement of the object, when monitoring the system for sufficiently long times. From our experimental data, we extract confidence levels of $90$\%, well beyond previous optical experiments. Our experiment also realizes a long-standing proposal for indirect Zeno measurements~\cite{Luis1996,Rehacek2000}. It presents the first IFM with a many-particle probe and opens the field of counterfactual quantum information~\cite{Jozsa1999,Hosten2006,Noh2009,Liu2012,Cao2014} to atom optics. Moreover, our setup opens the possibility to investigate open-system dynamics in the Zeno and anti-Zeno regime~\cite{Kofman2000,Facchi2001}.

\section{Results}
\textbf{Bose-Einstein condensate in an unstable spin configuration.} In our experiments, a \Rb BEC is prepared in the Zeeman level $\ketF{F, m_F}=\ketF{1,0}$, which is initially stable at a finite magnetic field (see Fig.~1~b). However, the Zeeman level $\ketF{1,-1}$ can be shifted by a microwave dressing on the transition to $\ketF{2,-2}$, until a resonance condition is reached~\cite{Klempt2009,Scherer2010}, and it becomes energetically favorable to populate the states $\ketF{1,\pm 1}$ by the decay of atom pairs. In this case, the BEC in $\ketF{1,0}$ parametrically amplifies quantum fluctuations in the levels $\ketF{1,\pm 1}$~\cite{Klempt2010}. After a given evolution time $t$, the output state is the so-called two-mode squeezed vacuum state~\cite{Gerry2005}:

\begin{equation}
	\ket{\xi}=\sum_{n=0}^{\infty} \frac{(-i \tanh \xi)^n}{\cosh{\xi}} \ket{n}_{-1} \ket{n}_{+1} .
\end{equation}

 Here, $\ket{n}_{\pm 1}$ denotes a Fock state of $n$ atoms in the level $\ketF{1,\pm 1}$. The state is characterized by the squeezing parameter $\xi = \Omega t$, where $\hbar \Omega$ describes the energy scale of the spin changing collisions. Most importantly, the unstable BEC generates an exponential increase of the number of atoms in the two levels $\ketF{1,\pm 1}$~\cite{Rehacek2000,Lamacraft2007} according to

\begin{equation}
	\bra{\xi} n_{\pm 1}\ket{\xi}=\sinh^2{\xi}
\end{equation}

Figure~1~c shows this exponential increase, which is used to determine the spin dynamics rate $\Omega=2 \pi \times 3.1$~s$^{-1}$.

\begin{figure}[t!]
		\includegraphics{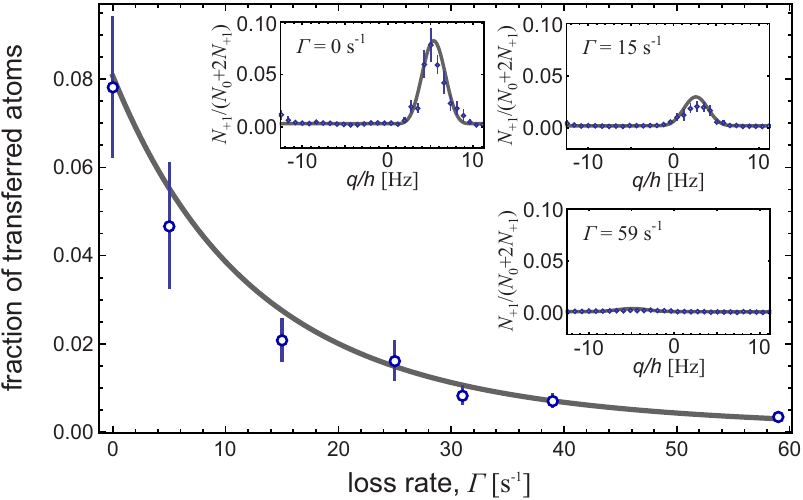}
		\caption{\textbf{Zeno suppression of quantum phase transition.} The number of atoms in the state $\ketF{1,1}$ produced by spin dynamics during $200$~ms (open circles) as a function of the effective loss rate $\Gamma$. The error bars indicate the statistical uncertainty due to the finite number of measurements. The insets present the spin dynamics resonances as a function of the energy difference $q$ between the input and the output states. The resonances are shown for the unperturbed case and for finite effective loss rates $\Gamma = 15\,\mathrm{s}^{-1}$ and $\Gamma = 59\,\mathrm{s}^{-1}$. The strong suppression of spin dynamics for an increased effective loss rate is well reproduced by a theoretical model without free parameters (grey lines). The error bar presents the standard error of the mean fraction of transferred atoms.}
	\label{fig2}
\end{figure}

\textbf{Zeno suppression of the decay.} At short evolution times, the unstable BEC features a quadratic increase of the probability of finding a single atom pair in the levels $\ketF{1,\pm 1}$, which is a prerequisite for the appearance of the quantum Zeno effect. While the quantum Zeno effect only refers to the decay of a single particle, our setup features a strong amplification of the signal. In the absence of the quantum Zeno effect, the unstable gas parametrically amplifies the single-atom-pair probability to a $1,000$-particle signal. Since the atoms are always transferred in pairs~\cite{Luecke2011,Luecke2014}, an "object" absorbing atoms in \textit{one} of the two output levels will suppress the decay to \textit{both} output levels.

\textbf{Indirect Zeno measurement.} In our experiments, the absorbing object is implemented by a laser beam which is resonant with the $F=2$ hyperfine state~\cite{Streed2006a,Facchi2009} (see Fig.~1). In combination with the microwave dressing, this laser beam generates an effective loss rate $\Gamma$ for the level $\ketF{1,-1}$ which can be freely controlled by the laser intensity. Figure~2 shows the effect of this loss on the spin dynamics instability. It demonstrates that the loss rate on the level $\ketF{1,-1}$ hinders and finally prevents the generation of atoms in the level $\ketF{1,1}$, although this level is not influenced directly. The experimental data is well reproduced by a master equation describing spin dynamics and the additional loss term (see Methods). Interestingly, our setup is equivalent to the proposal by Luis and Pe\v{r}ina which was initially devised, but never realized, for optical parametric down-conversion~\cite{Luis1996,Rehacek2000}. Furthermore, the atoms in $\ketF{1,-1}$ can be regarded as a decay product of the atoms decaying from $\ketF{1,0}$ to $\ketF{1,1}$. Since the Zeno measurement is performed on a decay product, the measurement is considered to be \textit{indirect}. In this sense, our results represent the first observation of the quantum Zeno effect with a continuous, indirect, negative-result measurement, 
which is regarded as the most stringent demonstration by some authors~\cite{Koshino2005}.

\begin{figure}[t!]
		\includegraphics{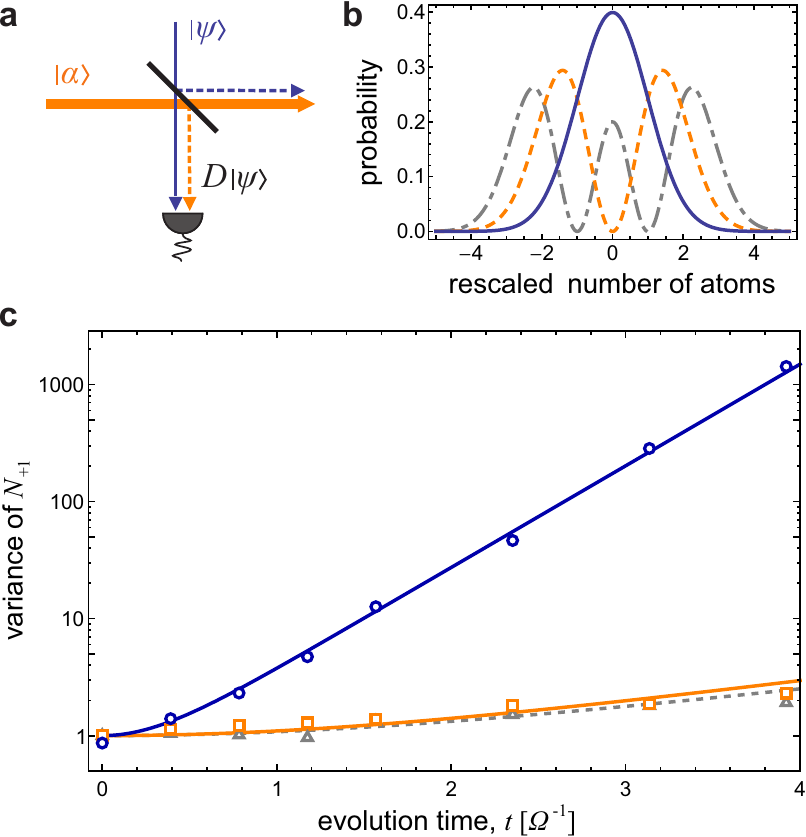}
		\caption{\textbf{Unbalanced homodyne detection. a} The weak coupling of a large coherent state $\ket \alpha$ with an arbitrary state $\ket \psi$ on an unbalanced beam splitter can be described by a displacement of the state, $D_\alpha \ket \psi$. In our experiment, the displacement is realized by a microwave coupling between the condensate in the level $\ketF{1,0}$ and the state in the level $\ketF{1,1}$. \textbf{b} The displacement of Fock states $\ket n$ results in characteristic particle counting statistics of the displaced states. The corresponding probability distributions, as shown for Fock states with $n=0$ (solid line), $1$ (dashed) and $2$ (dash-dotted) particles, are Hermite polynomials with a linearly increasing variance. \textbf{c} Variance of the rescaled number of atoms $N_{+1}$ in $\ketF{1,1}$ after unbalanced homodyne detection as a function of the evolution time. The variance in the absence of an absorbing object (blue circles) increases strongly with time according to Var$(D_\alpha \ket\xi)= \cosh(2\xi) V_\mathrm{sn}$ with $\xi = t\, \Omega$ and $\Omega = 2 \pi \times 3.1$~s$^{-1}$ (blue line). In the presence of an absorbing object (orange squares), the variance is almost constant. A slight increase is compatible with the increased variance after a holding time without spin dynamics (gray triangles) and can be attributed to residual rf noise. The solid orange and the dashed grey line are guides to the eye. All error bars are on the order of the symbol size.}
	\label{fig3}
\end{figure}

\textbf{Homodyne detection.} These measurements show that an object absorbing atoms in the level $\ketF{1,-1}$ can be detected, since it prevents the BEC's decay to the levels $\ketF{1,\pm 1}$. However, a proof of an IFM requires a detection of the object without the decay of a \textit{single atom} to the level $\ketF{1,-1}$. Such a proof presents a considerable challenge due to our atom counting uncertainty of $15$~atoms. We overcome this challenge by employing an unbalanced homodyne detection of the number of atoms in the level $\ketF{1,1}$~\cite{Gross2011}. It is known from quantum optics that a small coupling of a large coherent state $\ket \alpha$ with a small quantum state can be described by the action of a displacement operator $D_\alpha$~\cite{Paris1996} (see Fig.~3~a and b). While the vacuum state is displaced to a coherent state with a shot noise variance $V_\textrm{sn}$, the displaced nonzero Fock states exhibit a quickly increasing variance Var$(D_\alpha \ket{n})= (2 n+1) V_\textrm{sn}$. Due to these large differences, the contributions of the few-particle Fock states can be resolved after a displacement of the initial state. Hence, the homodyning technique is ideally suited to discriminate between unwanted few-particle measurements and clean zero-particle IFMs.

Experimentally, we implement the homodyning technique by using a short microwave pulse to couple the BEC with the atoms in the level $\ketF{1,1}$. Figure~3~c presents the measured variances after homodyning with and without the absorbing object. Without the object, the variance nicely follows the prediction  Var$(D_\alpha \ket{\xi})= \cosh{(2 \xi)} V_\textrm{sn}$ (see Methods). With the object, the variance is almost constant and does not show the drastic exponential increase. The large disparity of the underlying distributions allows for a reliable detection of the absorbing object.

\textbf{Application for interaction-free measurements.} It remains to be confirmed that this measurement is indeed interaction-free, which is achieved only if no atoms are transferred. In our case, the "with object"-variance is well below $3$, which would be the result of a displaced single-particle Fock state $\ket{1}$. This indicates that predominantly the displaced vacuum state is observed. In the following, we analyse the distributions after displacement for a squeezing parameter $\xi=3.1$ to extract the efficiency of our IFM.

\begin{figure}[t!]
		\includegraphics{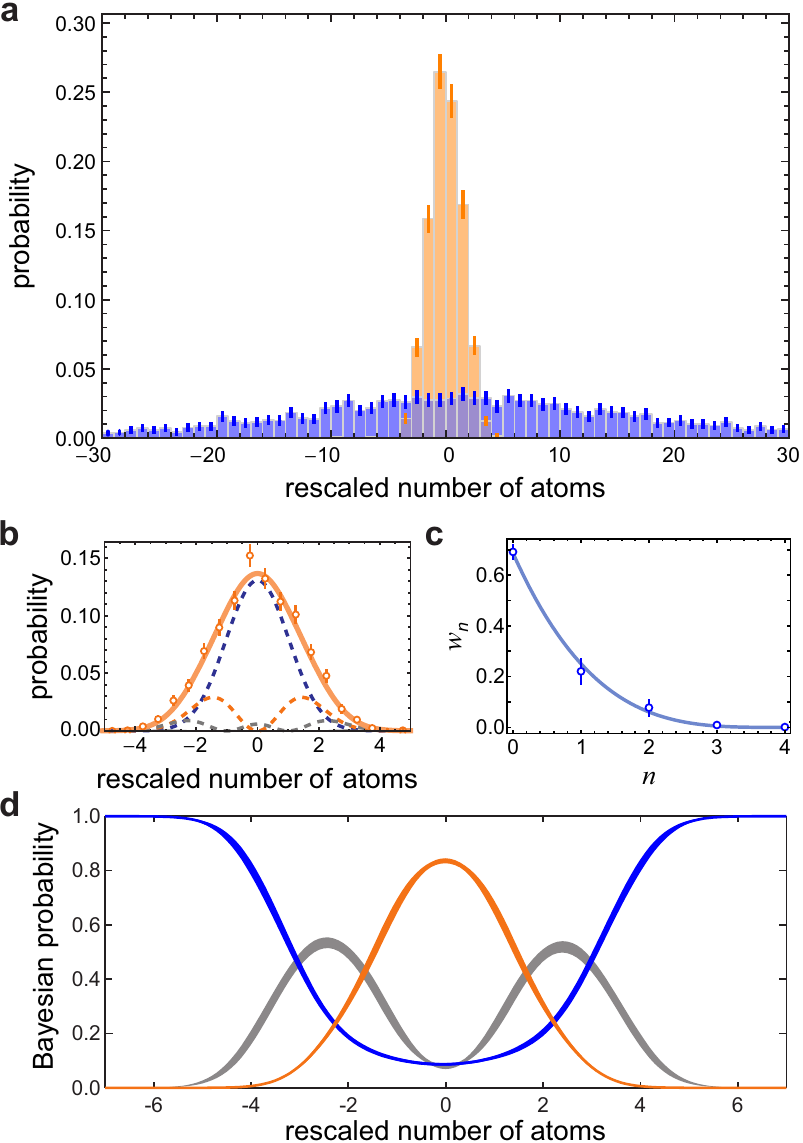}
		\caption{\textbf{Analysis of the interaction-free measurements. a} After the unbalanced homodyning, the counting statistics in the presence (orange bars) and absence (blue bars) of the Zeno suppression are very different and may be used to distinguish the "with object" and "without object" case. The error bars indicate the expected statistical noise for the finite set of measurements. \textbf{b} The counting statistics in the "with object" case (orange circles) can be reproduced by a weighted sum (solid orange line) of Hermite polynomials (dashed lines) corresponding to the Fock states $\ket n$ with $n=0,1,2,3$ and $4$. \textbf{c} Based on a Maximum Likelihood analysis, we can determine the weights $w_n$ of the individual Fock states. Most importantly, the vacuum contribution of $67(2)$\% shows that most measurements are indeed interaction-free. \textbf{d} Based on a given \textit{single} measurement result (corresponding to the x-axis), the probabilities for the three possible cases are given: "with object" (orange line), "without object" (blue line) and the case where an interaction has occurred (grey line). These results were obtained from a Bayesian analysis with an uninformative prior. The lines' thicknesses correspond to the uncertainties of these probabilities. The uncertainties in b), c), and d) are derived from the expected statistical uncertainties in a) by a bootstrapping method~\cite{Lvovsky2001}.}
	\label{fig4}
\end{figure}

Figure~4~a shows the result of $4,200$ homodyning measurements in the level $\ketF{1,1}$. The measured number of atoms has been rescaled such that the displaced vacuum state yields a normal distribution of width 1 centered at zero (see Methods). The resulting histograms are thus independent of the total number of particles in the specific realization of the homodyning measurement. These histograms reflect the underlying probability distributions for the outcome of a single measurement. The "with object"-distribution is analyzed further in Fig.~4~b. A Maximum Likelihood analysis (see Methods) allows for a reconstruction of the underlying state. The corresponding contributions are displayed in Fig.~4~c. Clearly, the vacuum state has the strongest weight with a contribution of $w_0=67(2)$\%, reflecting the interaction-free character of our measurements.

\section{Discussion}
The probability distributions obtained with and without object provide a calibration of our apparatus. They can be used to detect the presence/absence of an object from a single measurement result, without prior knowledge. For example, a rescaled atom number of $20$ is not compatible with the existence of an object, while a sufficiently small rescaled atom number strongly suggests the presence of an object. For an optimal discrimination between the two outcomes, the absolute value of the rescaled number of atoms should be compared to a threshold of $1.7$ (see Supplementary Note~1). In presence (absence) of the object, we obtain a measurement below (above) this threshold with a probability larger than $90$\%.

We evaluate the figure of merit introduced in the original proposal~\cite{Elitzur1993} $\eta = \frac{ P(\mathrm{D}) }{ P(\mathrm{D}) + P(\mathrm{int})}$, where $P(\mathrm{D})$ is the probability of performing an IFM and $P(\mathrm{int})$ is the probability of interaction with the object. In the case with object, this parameter yields the optimal probability of detecting it without interaction. We generalize the parameter for the experimentally relevant case that an object can only be detected with a finite confidence~\cite{Kwiat1999}. From our measurements, we obtain a figure of merit of $\eta=65(2)$\% at our confidence level of $90$\% (see Supplementary Note 1). It exceeds the threshold of the original Elitzur-Vaidman scheme~\cite{Elitzur1993} and reaches a value comparable to the one achieved in optics experiments, although at a much higher confidence level ($\eta=63(1)$\% in the experiment of Ref.~\citenum{Kwiat1999} with a confidence of $65$\%, $\eta \approx 50$\% in Ref.~\citenum{Kwiat1995a} with a confidence of $2$\%). An improvement of the mentioned optics results with state-of-the-art technology is to be expected but outstanding.

The confidence of possible statements for any measurement result can be determined by a Bayesian analysis (see Methods) of the recorded probability distribution (Fig.~4~d). The results show that the existence of an object can be inferred interaction-free with a confidence level of up to $84(1)$\% at a rescaled number of atoms of $0$. The absence of an object can be detected with almost $100$\% confidence for a wide range of possible measurement results beyond $\pm 6$. It is also possible to extract the probability for an interaction with the object. In principle, the interacting case can also be detected externally, for example by measuring the fluorescence photons scattered by the unwanted atoms in the level $\ketF{1,-1}$. If such an external detection was realized, the Bayesian confidence for an interaction-free detection of the object could be increased to $90.6(6)$\%.

Our analysis demonstrates that a BEC in an unstable spin configuration can be used for highly efficient IFMs. While the Zeno effect has already been demonstrated with BECs in previous publications~\cite{Streed2006a,Schafer2014},  a proof of an IFM requires a Zeno suppression and its detection on the single-atom scale which has not been presented up to now. Moreover, we emphasize that there are two versions of IFMs~\cite{Elitzur1993} depending on whether a possible interaction with the object can also be detected from the measurement result or not. In previous proposals~\cite{Elitzur1993,Vaidman2003,Kwiat1995a,Kwiat1999,Paul1996,Tsegaye1998}, such a detection requires single-particle sources. Our protocol realizes the strong version of IFMs without requiring single-particle sources. The achieved figure of merit ($65$\%) surpasses the predicted optimum of the original proposal ($50$\%) and is comparable to the best results obtained in optics experiments which also rely on the quantum Zeno effect~\cite{Kwiat1999}, yet with a larger confidence of $90$\%. For improved atom counting and a noise-less environment, the method permits a $100$\% figure of merit with an almost ideal confidence of $1 - 4/\exp(2 \xi)$. 

\section{Methods}
\begin{scriptsize}

\textbf{\footnotesize Initial experimental sequence.}
We start the experiments with an almost pure Bose-Einstein condensate of $25,000$ $^{87}$Rb atoms in an optical dipole potential with trap frequencies of $2 \pi \times (200,150,150)$~Hz. At a homogeneous magnetic field of $2.6$~G ($70 \mu$G), the condensate is transferred to the state $\ketF{1,0}$ is prepared by a series of three resonant microwave pulses. During this preparation, two laser pulses resonant to the $F=2$ manifold purify the system from atoms in unwanted spin states. Directly before spin dynamics is initiated, the output states $\ketF{1,\pm 1}$ are emptied with a pair of microwave $\pi$-pulses from $\ketF{1,+1}$ to $\ketF{2,+2}$ and from $\ketF{1,-1}$ to $\ketF{2,-2}$ followed by another light pulse. The lifetime of the condensate in the state $\ketF{1,0}$ is $19$~s as a result of background gas collisions and three-particle loss.

\textbf{\footnotesize Theoretical description of the Zeno effect.}
Our scheme for an interaction-free measurement relies on an application of the quantum Zeno effect in an unstable spinor Bose-Einstein condensate. In this section, we extend our description of a spinor condensate in an unstable spin configuration~\cite{Klempt2009,Klempt2010,Deuretzbacher2010,Scherer2010,Scherer2013} by an additional Zeno measurement of one of the output states. 

The Hamiltonian describing pair creation in $m = \pm 1$ in a single spatial mode due to spin-changing collisions 
reads~\cite{Klempt2009,Klempt2010,Deuretzbacher2010,Scherer2010,Scherer2013}:
\begin{equation}
\hat H = (\epsilon + q) \left( \hat a_{-1}^\dagger \hat a_{-1} + \hat a_{+1}^\dagger \hat a_{+1} \right) + \Omega \left( \hat a_{-1}^\dagger \hat a_{+1}^\dagger 
+ {\mathrm H.c.}\right),
\end{equation}
where the operators $\hat a_{\pm 1}^\dag$ create particles in $m=\pm 1$, $\epsilon$ is the energy of the resonant mode, 
$q$ is the quadratic Zeeman energy, and $\Omega = U_1 \int d \mathbf{r} \, n_\text{BEC} (\mathbf{r}) |\phi (\mathbf{r})|^2$ is the strength of the pair creation, with  
$n_\text{BEC}$ the density of the condensate in $m=0$, $\phi$ the wave function of the resonant mode, 
$U_1 = (g_2-g_0)/3$, and $g_F = 4 \pi \hbar^2 a_F / M$~($M$ is the atomic mass and $a_F$ the $s$-wave scattering length for the collisional channel with total spin $F$). 

In the presence of losses in $m=-1$ with a loss rate $\Gamma$, the dynamics of the density operator $\hat\rho$ is given by the Lindblad master equation:
\begin{equation}
\frac{d \hat\rho}{dt} = -\frac{i}{\hbar} \left[ \hat H , \hat\rho \right] + \frac{\Gamma}{2} \left( 2 \hat a_{-1} \hat \rho \hat a_{-1}^\dagger - 
\hat a_{-1}^\dagger \hat a_{-1} \hat \rho - 
\hat \rho \hat a_{-1}^\dagger \hat a_{-1} \right).
\end{equation}
using this master equation we may evaluate the time evolution of the average of any operator $\hat O$, 
$\frac{d}{dt}\langle \hat O \rangle=\mathrm{Tr}\{\hat O \frac{d\hat\rho}{dt} \}$. 
Defining the populations $N_{-1} \equiv \langle \hat a_{-1}^\dagger \hat a_{-1} \rangle$, $N_{+1} \equiv \langle \hat a_{+1}^\dagger \hat a_{+1} \rangle$, 
and the pair correlations $2 {\mathrm i} v \equiv \langle \hat a_{-1}^\dagger \hat a_{+1}^\dagger - \hat a_{-1} \hat a_{+1} \rangle$, 
$2u \equiv \langle \hat a_{-1}^\dagger \hat a_{+1}^\dagger + \hat a_{-1} \hat a_{+1} \rangle$, we obtain the coupled Bloch-like equations of motion:
\begin{eqnarray}
\frac{dN_{-1}}{dt}  & = & - \Gamma N_{-1}+ \frac{2 \Omega}{\hbar} v, \\
\frac{dN_{+1}}{dt} & = & \frac{2 \Omega}{\hbar} v, \\
\frac{dv}{dt} & = & - \frac{\Gamma}{2} v + \frac{\Omega}{\hbar} (1 + N_{-1} + N_{+1}) + \frac{2 (\epsilon + q)}{\hbar} u, \\
\frac{du}{dt} & = & - \frac{\Gamma}{2} u - \frac{2 (\epsilon + q)}{\hbar} v,
\end{eqnarray}
which we solved numerically for $N_{\pm 1} = 0$ at $t=0$. The comparison between the results of this coupled system and our experimental results~(Fig.~2) 
require an independent calibration of the Zeno measurement rate $\Gamma$ and the spin dynamics rate $\Omega$. We measure a spin dynamics rate 
of $\Omega=2\pi\,3.6\,\mathrm{s}^{-1}$ for the data set of Fig.~2 due to a slightly different setup compared to the measurements 
of the main results of the paper.

\textbf{\footnotesize Calibration of the Zeno measurement rate.}
An object absorbing atoms in the state $\ketF{1,-1}$ suppresses the decay of the BEC in the state $\ketF{1,0}$.  This suppression can be described as a continuous Zeno measurement of the number of atoms in the state $\ketF{1,-1}$, where the measurement rate corresponds to the absorption rate of the object. In our experiments, the absorbing object is implemented by a resonant laser beam on the $F=2$ hyperfine manifold, which expels atoms from the trap. Since the state $\ketF{1,-1}$ is coupled to the state $\ketF{2,-2}$ by a weak microwave dressing field, the combined microwave and optical fields result in an effective loss rate for atoms in the state $\ketF{1,-1}$. We calibrate this effective loss rate by preparing a sample of atoms in the state $\ketF{1,-1}$. While microwave and laser light are switched on, we record the number of remaining atoms as a function of exposure time. Supplementary Fig.~1 shows the experimental results of such a measurement. For our experimental parameters, the loss follows an exponential decay. We extract the effective loss rate $\Gamma$ from exponential fits to the data.

The optical coupling of the resonant laser beam also leads to a small shift of the resonance position, as can be seen in Fig.~2 (insets). The shift depends directly on the intensity of the resonant laser and on the corresponding effective loss rate. Supplementary Fig.~2 shows the measured resonance position as a function of the effective loss rate. For large effective loss rates, as desired for the interaction-free measurements, the strong Zeno suppression prevents a simple measurement of the resonance position. In these cases, we enhance the transfer of atoms to the states $\ketF{1,\pm 1}$ by generating a seed population in the state $\ketF{1,1}$. These calibration measurements ensure that our data is always taken on the spin dynamics resonance with an independently recorded effective loss rate $\Gamma$.

\begin{figure}[ht!]
		\includegraphics{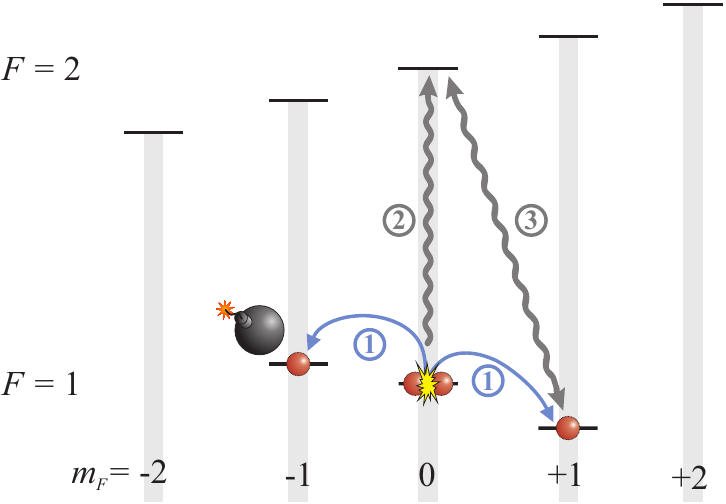}
		\caption{\textbf{\footnotesize Realization of the atomic homodyning.} After an evolution the in the presence or absence of an absorbing object in the $\ketF{1,-1}$ state (1), we transfer the remaining atoms in the $\ketF{1,0}$ state to $\ketF{2,0}$ with a microwave pulse (2). These atoms act as a strong coherent state for the displacement of the state in $\ketF{1,-1}$ by a short microwave pulse (3).}
	\label{fig5}
\end{figure}

\textbf{\footnotesize Atomic homodyning.}
In our experiments, an interaction-free measurement is only achieved if not a single atom is transferred to the $\ketF{1,-1}$ state. Hence, an atom detection on the single particle level is necessary. We solve this technical challenge by implementing an unbalanced homodyne detection for atoms as described in the following.

The method is equivalent to a displacement of the state created in $\ketF{1,+1}$. The particle counting statistics of displaced Fock states can be described by Hermite polynomials~\cite{Lvovsky2009} as depicted in Fig.~3~b. The variance of these polynomials Var$(D\ket n)=(2n+1)\,V_\mathrm{sn}$ is proportional to the number of particles $n$. Here, $V_\mathrm{sn}$ is the shot noise of the displaced vacuum Var$(D\ket 0)=V_\mathrm{sn}$. Hence, for the displaced squeezed state we expect a variance of
\begin{eqnarray}
\mathrm{Var}(D\ket{\xi}) &=&
(2 \bra{\xi}n\ket{\xi} + 1)\, V_\mathrm{sn}\nonumber\\
&=& (2 \sinh^2(\xi) + 1)\, V_\mathrm{sn}\nonumber\\
&=& \cosh(2\xi)\, V_\mathrm{sn}
\end{eqnarray}
To extract the Fock-state contributions, we can reproduce the recorded histograms by a weighted sum of these polynomials (see Fig.~4~b). The weights are gained from a Maximum Likelihood analysis~\cite{Banaszek1998}. Moreover, we convolve the theoretical distributions of the displaced Fock states with a Gaussian distribution to account for our detection noise. This affects our results only slightly and thus shows that the detection noise does not corrupt the homodyning technique. We have checked that our results are consistent with the Fock state contributions calculated by the pattern-function method~\cite{Lvovsky2009}.

For the implementation of the unbalanced homodyne detection in our experiments, the remaining condensate in $\ketF{1,0}$ can be used as the strong coherent state. Since all particle numbers are measured in the end, the state is indeed closer to a Fock state. We have checked however, that at these large particle numbers, the homodying results for Fock and coherent state are equivalent. For the realization of the unbalanced beam splitter, we first transfer these atoms to the $\ketF{2,0}$ level and then apply a short microwave pulse to couple the coherent state to the state created in the $\ketF{1,+1}$ level (see Fig.~\ref{fig5}). If no atoms are present in the $\ketF{1,+1}$ level, this pulse transfers about $\cos^2\theta=8$\% of the condensate, where $\theta=\omega t$ with the microwave Rabi frequency $\omega$ and the pulse duration $t$.

However, the shot-to-shot variation of the number of atoms $N_0$ in the condensate leads to a fluctuating number of transferred atoms. To compensate for these fluctuations we subtract $\cos^2\theta \;(N_0'+N_{+1}')$ from the measured number of atoms in the $\ketF{1,+1}$ level such that the resulting distribution is always centred at 0 regardless of the total number of atoms $(N_0'+N_{+1}')$. Here $N_0'$ and $N_{+1}'$ are the number of atoms measured after the displacement in the corresponding states. Additionally, we rescale the number of particles such that the variance of an ideal displaced vacuum state is $V_\mathrm{sn}=1$ and obtain
\begin{equation}
\tilde N_{+1} =
\frac {N_{+1} - \cos^2\theta \; (N_0'+N_{+1}')}
	{\sqrt{\cos^2\theta\;(1-\cos^2\theta)(N_0' + N_{+1}')}}
\end{equation}
The resulting distribution is further analyzed to extract the Fock state contributions as described in the previous section.

Our detection system was calibrated for correct counting of up to $6,000$ atoms in a single cloud using two independent methods~\cite{Luecke2011}. However, for the described rescaling method it is essential to correctly measure numbers of particles up to $25,000$ atoms in the coherent state. In this regime, we have to correct for a slight non-linearity of our detection due to the finite imaging resolution. This effect was independently measured by comparing the measured number in a complete cloud with a cloud that was separated in two Zeeman levels. The nonlinearity is negligible for up to $10,000$ atoms in a single cloud. It increases for larger numbers, up to a value of $15$\% for the large coherent state. In spite of our detection noise of 16 atoms, the described homodyning technique allows for statistical statements about the number of particles on a single atom level. This technique is thus essential for the claim of interaction-free measurements. Moreover, we believe that this technique opens the door for new experiments which require measurements on the single particle level.

\textbf{\footnotesize Bayesian analysis.}
The Elitzur-Vaidman figure of merit $\eta$ considers only the case with object. To qualify how well the interaction-free measurement discriminates between the presence and the absence of the object after a single measurement, we employ a Bayesian analysis.

Let $P(\num \vert \y)$ and $P(\num \vert \n)$ be
the counting statistics of rescaled atom number in presence 
($\y$ -- ''with object'' case) and absence 
($\n$ -- ''without object'' case) of Zeno dynamics, respectively.
The counting statistics in the $\y$ case is decomposed 
as the weighted sum 
\be \label{Pny}
P(\num \vert \y) = \sum_{n \geq 0} 
P(\num \vert n, \y) \, \weightn
\ee
of Hermite polynomials.
The function $P(\num \vert n, \y)$ is
the probability to detect a rescaled atom 
number $\num$ after homodyne, if a Fock state $\vert n \rangle$ was present in $m_F=+1$ before homodyne. 
The coefficients $\weightn \equiv P(n \vert \y)$ 
are equivalent to those shown in Fig.~4~c and correspond to the 
probability to have $n$ particles at the end 
of the Zeno dynamics.
Let $p$ be the probability that the object is present. The overall counting statistics is 
\be \label{Pn}
P(\num \vert p) = p P(\num \vert \y) 
+ (1-p) P(\num \vert \n). \nonumber
\ee
We use the Bayes's theorem
[$P(X\vert Y)P(Y) = P(Y\vert X)P(X)$, 
$X$ and $Y$ being stochastic variables]
to calculate the conditional probabilities:
\be
P({\rm NO} \vert \num, p) = 
\frac{ (1-p) P(\num \vert \n) }{ P(\num \vert p) } 
\ee
that the object is not there; 
\be \label{Pinf}
P({\rm IFM} \vert \num, p) = 
\frac{ p \,\weighto \, P(\num \vert 0, \y) }{ P(\num \vert p) }
\ee
that the object is there and no interaction has occured, 
corresponding to an interaction-free measurement (IFM) event; and
\be
P({\rm int} \vert \num, p) = 
\frac{ p \sum_{n > 0} \weightn P(\num \vert n, \y) }
{ P(\num \vert p) }
\ee
that the object is there and an interaction has occured. These probabilities correspond to the dashed blue, solid orange, and dotted grey lines in Fig.~4~d, respectively, calculated in the uninformative prior condition $p=1/2$.
\end{scriptsize}

\section{Acknowledgments}
\begin{scriptsize} We thank H. Bachor, E. Rasel and S. Pascazio for inspiring discussions. We acknowledge support from the Centre for Quantum Engineering and Space-Time Research QUEST and from the Deutsche Forschungsgemeinschaft (Research Training Group 1729 and project SA 1031/7-1). We also thank the Danish Council for Independent Research, and the Lundbeck Foundation for support. We acknowledge support from the European Metrology Research Programme (EMRP). The EMRP is jointly funded by the EMRP participating countries within EURAMET and the European Union. 
\end{scriptsize}

\newpage
\clearpage
\setcounter{figure}{0}

\renewcommand{\figurename}{\textbf{Supplementary Figure}}

\begin{figure}[ht!]
		\includegraphics[width={86mm}]{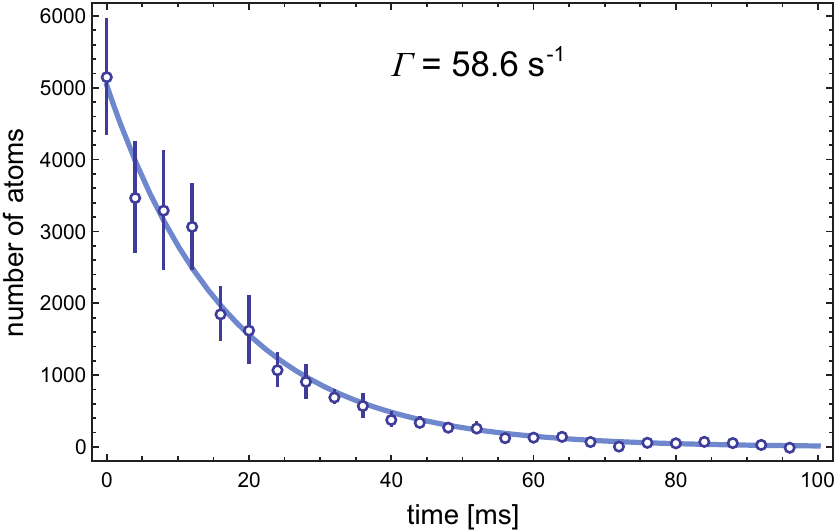}
		\caption{\textbf{Calibration of the loss rate.} Mean number of particles in the state $\ketF{1,-1}$ versus the exposure time. For this exemplary light intensity an exponential fit to the data (blue solid curve) yields a loss rate of $\Gamma=58.6\,\mathrm{s}^{-1}$.}
	\label{figS1}
\end{figure}

\begin{figure}[ht!]
		\includegraphics[width={86mm}]{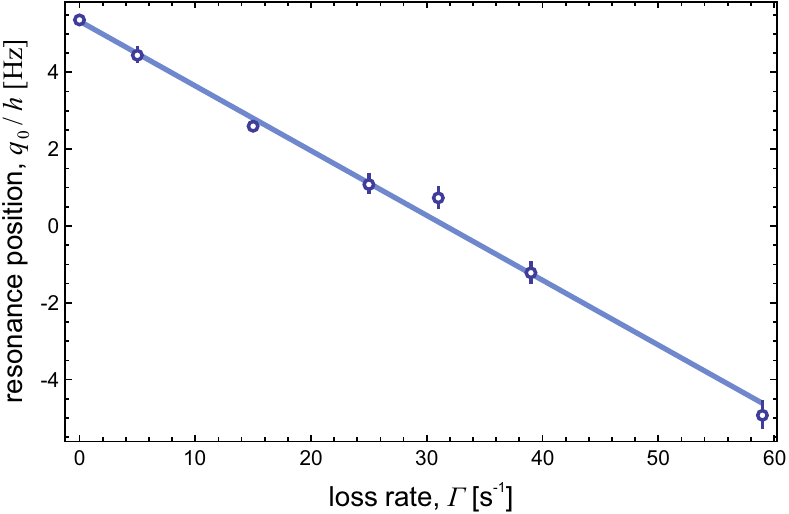}
		\caption{\textbf{Resonance position versus loss rate.} The position of the spin dynamics resonance is shifted depending on the loss rate $\Gamma$. The solid blue line is a linear fit to the data.}
	\label{figS2}
\end{figure}

\begin{figure}[ht!]
		\includegraphics[width={80mm}]{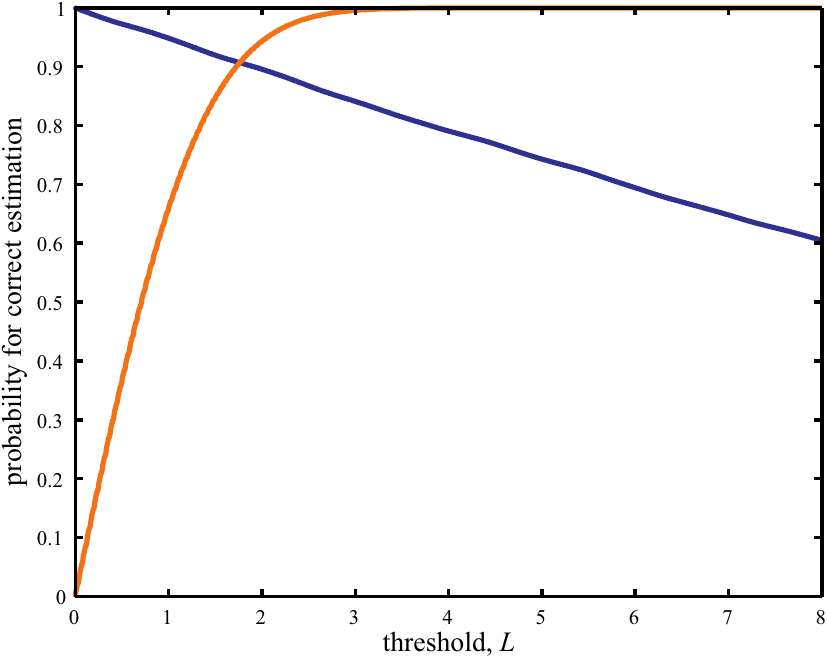}
		\caption{\textbf{Discrimination probabilities as a function of the threshold $L$.} In the "with object" ("without object") case, the orange (blue) line represents the probability that a single measurement result lies inside (outside) the range from $-L$ to $L$.}
	\label{figS4}
\end{figure}

\hfill
\clearpage
\newpage
\section*{Supplementary Note 1: Elitzur-Vaidman figure of merit}
In this section, we derive a figure of merit for the demonstrated interaction-free measurements in analogy to the original Elitzur-Vaidman (EV) proposal~[2]. In the EV proposal, the measurement is repeated until the photon either interacts with the object or exits the dark port $D$ (see Fig.1a). 
The probability of an interaction-free measurement is thus given by
\begin{eqnarray}
\eta &=& P(\mathrm{D}) + P(\mathrm{B})P(\mathrm{D}) + P(\mathrm{B})^2 P(\mathrm{D}) + ... \nonumber  \\ &=& P(\mathrm{D}) \big[ 1 + P(\mathrm{B}) + P(\mathrm{B})^2 + ... \big]  = \frac{P(\mathrm{D})}{1-P(\mathrm{B})}, \nonumber
\end{eqnarray}
where we summed the geometric series after an infinite number of trials. Taking into account that $P(\mathrm{D})+P(\mathrm{B})+P({\rm int}) = 1$, we recover the familiar figure of merit
\begin{equation}
\eta = \frac{P(\mathrm{D})}{P(\mathrm{D})+P({\mathrm int})}. \nonumber
\end{equation}

We generalize the figure of merit for the experimentally relevant case that an object can only be detected with a finite confidence. In our case, it is necessary to define a threshold $L$ to discriminate between the presence and the absence of the object after a single measurement result. As in the original proposal, these measurement results are only evaluated if no interaction with the object took place. 

Supplementary Figure~\ref{figS4} shows the probability that a single measurement result lies inside the range from $-L$ to $L$ for the case with object (orange line). The probability increases monotonously from zero if $L=0$ to almost $100$\% for $L>4$. The blue line represents the probability that a single measurement result lies outside the range from $-L$ to $L$ for the case without object. It decreases slowly and almost linearly from $100$\% at $L=0$. At $L=1.7$, the two probabilities become equal: This threshold is optimal in the sense that both cases are treated symmetrically. For this threshold, we achieve a confidence of $90$\% for both the presence and the absence of the object.

The EV proposal considers three different outcomes for the case with object: (i) interaction with the object, (ii) interaction-free measurement of the object, and (iii) inconclusive outcome. These three results correspond to the following three outcomes in our experiments: (i) interaction with the object with a $33$\% probability, (ii) a measurement result within $-1.7$ to $1.7$, $60$\%, and (iii) a measurement result outside, $7$\%. The EV proposal requires a repetition of the measurement in case (iii). The figure of merit $\eta$ is then calculated as the probability to detect an existing object without interaction after a conclusive series of measurements. For our confidence of $90$\%, we obtain a corresponding figure of merit $\eta=65(2)$\%. Of course, this figure of merit could be further improved at the expense of a lower confidence for the "without object" case.

\end{document}